\documentclass[prd,showpacs,amsmath,amssymb,superscriptaddress,floatfix,nofootinbib,10pt]{revtex4}
\usepackage{times}
\usepackage{amssymb,amsbsy,amsmath,amsfonts}
\usepackage{graphicx}
\usepackage{float}
\usepackage{color}
\usepackage{morefloats}
\usepackage{rotating}
\usepackage{srcltx}
\usepackage{slashed}
\usepackage{multirow}
\usepackage{verbatim}
\usepackage{hyperref}
\usepackage{tabularx}

\newcommand{\PreserveBackslash}[1]{\let\temp=\\#1\let\\=\temp}
\newcolumntype{C}[1]{>{\PreserveBackslash\centering}p{#1}}
\newcolumntype{R}[1]{>{\PreserveBackslash\raggedleft}p{#1}}
\newcolumntype{L}[1]{>{\PreserveBackslash\raggedright}p{#1}}

\begin{document}

\title{$\Lambda_c N$ interaction in leading order covariant chiral effective field theory}

\author{Jing Song}
\affiliation{School of Physics, Beihang University, Beijing, 102206, China}

\author{Yang Xiao}
\affiliation{School of Physics, Beihang University, Beijing, 102206, China}
\affiliation{Universit\'e Paris-Saclay, CNRS/IN2P3, IJCLab, Orsay, 91405, France}

\author{Zhi-Wei Liu}
\affiliation{School of Physics, Beihang University, Beijing, 102206, China}

\author{Chun-Xuan Wang}
\affiliation{School of Physics, Beihang University, Beijing, 102206, China}

\author{Kai-Wen Li}
\email[E-mail me at: ]{kaiwen.li@buaa.edu.cn}
\affiliation{Beijing Advanced Innovation Center for Big Data-Based Precision Medicine, School of Medicine and Engineering, Beihang University, Key Laboratory of Big Data-Based Precision Medicine (Beihang University), Ministry of Industry and Information Technology, Beijing, 100191, China}
\affiliation{School of Physics, Beihang University, Beijing, 102206, China}

\author{Li-Sheng Geng}
\email[E-mail me at: ]{lisheng.geng@buaa.edu.cn}
\affiliation{School of Physics, Beihang University, Beijing, 102206, China}
\affiliation{Beijing Advanced Innovation Center for Big Data-Based Precision Medicine, School of Medicine and Engineering, Beihang University, Key Laboratory of Big Data-Based Precision Medicine (Beihang University), Ministry of Industry and Information Technology, Beijing, 100191, China}
\affiliation{Beijing Key Laboratory of Advanced Nuclear Materials and Physics, Beihang University, Beijing, 102206, China}
\affiliation{School of Physics and Microelectronics, Zhengzhou University, Zhengzhou, Henan, 450001, China}

\begin{abstract}

We study the  $\Lambda_c N$ interaction in the covariant chiral  effective field theory (ChEFT) at leading order. All the relevant low-energy constants are determined by fitting to the lattice QCD simulations from the HAL QCD Collaboration. Extrapolating the results to the physical point, we show that the $\Lambda_c N$ interaction is weakly attractive in the $^1S_0$ channel, but in the $^3S_1$ channel, it is only attractive at extremely low energies and soon turns repulsive for larger laboratory energy. Furthermore, we show that the neglect of the  $^3S_1-{}^3D_1$ coupling provided by the leading order covariant ChEFT would result in an attractive interaction in the $^3S_1$ channel at the physical point, which coincides with the previous non-relatistic ChEFT study. As a byproduct, we predict the $^3D_1$ phase shifts and the mixing angel $\varepsilon_1$, which can be checked by future lattice QCD simulations. In addition, we compare the $\Lambda_c N$ interaction with the $\Lambda N$ and $NN$ interactions to study how the baryon-nucleon ($BN$) interactions evolve as a function of the baryon mass with the replacement of a light quark by a strange or  charm quark in the baryon ($B$).

\end{abstract}


\date{\today}

\maketitle
\section{Introduction}
Baryon-baryon ($BB$) interactions are one of the most important inputs in studies of  hadronic matter. At present, the low energy  nucleon-nucleon ($NN$) interaction has already been comprehensively studied both phenomenologically and model independently~\cite{Fleming:1999bs,Bedaque:2002mn,Epelbaum:2008ga,Machleidt:2011zz,Ren:2016jna}. The investigation of the hyperon-nucleon ($YN$) interaction has achieved significant success as well~\cite{Korpa:2001au,Haidenbauer:2007ra,Haidenbauer:2013oca,Polinder:2007mp,Haidenbauer:2015zqb,Haidenbauer:2009qn}. Hypernuclear spectroscopy provides one of the most important sources from which  $Y N$ and hyperon-hyperon ($Y Y$) interactions can be derived. As a natural extension of $NN$ and $YN$ interactions, the charmed hyperon-nucleon ($Y_c N$) interaction,  has also been studied with growing interests~\cite{Dover:1977jw,Bando:1981ti,Bando:1983yt,Gibson:1983zw,Bando:1985up,Liu:2011xc,Maeda:2015hxa,Froemel:2004ea}. 
High energy facilities such as BEPC in China~\cite{Nishida:2019yvf,Li:2019ikx}, J-PARC in Japan~\cite{Fujioka:2017gzp}, and FAIR in Germany~\cite{Carames:2018xek} all have ongoing/proposed experiments on charm physics, for instance the production of $\Lambda_c$ and $\Sigma_c$ hyperons and their interactions with other hadrons~\cite{PANDA:2020zwv}.

 Early theoretical studies, based on either meson-exchange models~\cite{Dover:1977jw,Bando:1981ti,Bando:1983yt,Gibson:1983zw,Bando:1985up,Liu:2011xc} or constituent quark models~\cite{Froemel:2004ea,Maeda:2015hxa}, indicated that the $Y_cN$ ($Y_c=\Lambda_c, \Sigma_c$) interaction is fairly attractive. Particularly, compared with the $\Lambda N$ interaction, the strange meson ($K, K^*$) exchanges are replaced by the charmed meson ($D, D^*$) exchanges in the $\Lambda_cN$ interaction~\cite{Bando:1981ti,Bando:1983yt,Bando:1985up} in meson-exchanged models. This  would  result in less (more) attraction in the $S$- ($P$-) partial waves in the $\Lambda_c N$ interaction than  in the corresponding $\Lambda N$ potential, because of the larger masses of exchanged mesons.

Recently, the HAL QCD Collaboration has performed lattice QCD simulations for unphysical pion masses to study the $\Lambda_cN$ interaction~\cite{Miyamoto:2017tjs}. They obtained the $S$-wave phase shifts for $m_\pi=410$ MeV, $570$ MeV, and $700$ MeV. These results were subsequently studied in the next-to-leading order non-relativistic chiral effective field theory (ChEFT) and extrapolated to the physical point and a moderately attractive $\
\Lambda_c N$ interaction was found for both the $^1S_0$ and $^3S_1$ channels~\cite{Haidenbauer:2017dua}. In a later work, the $S$-wave $\Sigma_c N$ interaction with isospin $1/2$ was studied by the HAL QCD Collaboration as well~\cite{Miyamoto:2017ynx}. Taking these results as inputs, Meng et al.~\cite{Meng:2019nzy} calculated the $\Sigma_cN$ interaction to the next-to-leading order in the non-relativistic ChEFT and found that the $^3S_1$ interaction for isospin $1/2$ is weakly attractive, but the interaction for isospin $3/2$  is strongly attractive, resulting in a $\Sigma_c N$ bound state. It should be stressed that the latter prediction depends on the quark model inputs adopted.

As all the lattice QCD simulations of the $Y_c N$ system were still performed for unphysical light quark (pion) masses, a reliable extrapolation of these results to the physical point is essential to guide future experiments and to gain insights on the $Y_c N$ interaction. In a series of recent works, we have shown that the recently proposed covariant ChEFT approach can be used for such a purpose. In Refs.~\cite{Li:2016mln,Song:2018qqm,Ren:2018xxd,Li:2018tbt}, it was shown that the strangeness $S=-1$ lattice QCD $YN$ interaction can be described reasonably well, so is the strangeness $S=-2$ $YN$/$YY$ interaction. The extrapolated results are also consistent with limited experimental data. In a more recent work~\cite{Bai:2020yml}, we showed that the lattice QCD $NN$ phase shifts for the $^1S_0$ and $^3S_1-{}^3D_1$ partial waves can be simultaneously described together with their physical counterparts by the leading order (LO) covariant ChEFT, implying that a reliable chiral extrapolation of lattice QCD results for unphysical pions masses smaller than 500 MeV is possible.  An interesting discovery of Ref.~\cite{Bai:2020yml} is that the $^3S_1-{}^3D_1$ coupled channel is described by the same two low-energy constants (LECs), thus allowing one to make predictions 
for the $^3D_1$ phase shifts and the mixing angle $\varepsilon_1$ using only the $^3S_1$ phase shifts as inputs.

In this work, we revisit the HAL QCD results~\cite{Miyamoto:2017tjs} in the covariant ChEFT  at leading order. Our purpose is threefold. First, we provide an independent extrapolation of the HAL QCD results, in addition to that of Ref.~\cite{Haidenbauer:2017dua}. Second, we predict the $^3D_1$ phase shifts and the mixing angle $\varepsilon_1$, so that they can be checked by future lattice QCD simulations, which could also provide a non-trivial check of the covariant ChEFT. Third, we investigate the quark mass dependence of baryon-nucleon interactions by comparing those of $NN$, $\Lambda N$, and $\Lambda_c N$. 

This paper is organized as follows. In Sec.~II, we briefly introduce the covariant ChEFT for the $Y_cN$ system, including coraviant chiral Lagrangians, potentials, and the scattering equation, as well as our strategy to determine the unknown LECs. In Sec.~III we show the fitted results and extrapolations, and discuss about coupled channel effects and quark mass dependence in the $NN$, $\Lambda N$, and $\Lambda_c N$ interactions. Followed by a short summary and outlook in Sec.~IV.

\section{Leading order covariant chiral effective field theory}
\begin{figure}[h]
  \centering
  \includegraphics[width=0.2\textwidth]{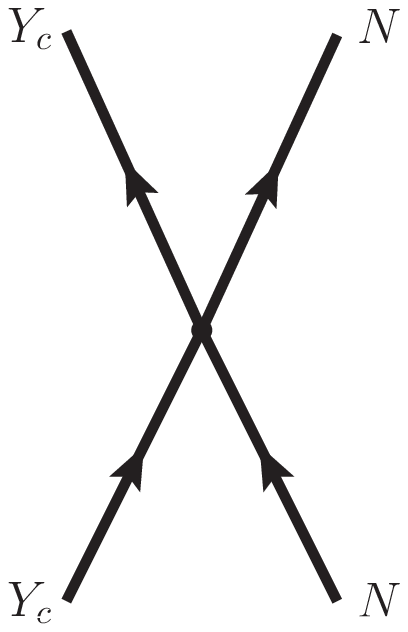} \qquad
  \includegraphics[width=0.2\textwidth]{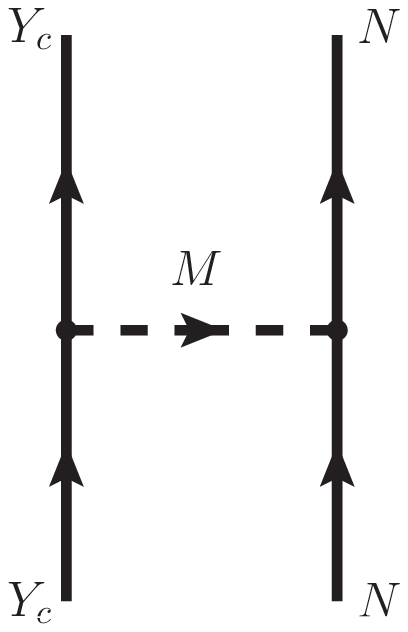}
  \caption{Leading order Feynman diagrams for non-derivative four-baryon contact terms and one-meson-exchanges.}\label{CTOME}
\end{figure}

In this section, we briefly introduce the covariant ChEFT for the $Y_cN$ interaction.  At leading order, the $Y_cN$ potentials consist of contributions from non-derivative four-baryon contact terms (CT) and one-meson-exchanges (OME), as shown in Fig.~\ref{CTOME}.

The LO Lagrangian for the contact terms is:
 \begin{align}\label{CT}
  \mathcal{L}_{\textrm{CT}}^{Y_cN\rightarrow Y_cN} = C_i~\left(\bar Y_c \Gamma_i Y_c) (\bar N \Gamma_i N\right),
\end{align}
where $C_i$ ($i=1\ldots 5$) are the LECs that need to be determined by fitting to either experimental or lattice QCD data, and $\Gamma_i$ ($i=1\ldots 5$) are the elements of the Clifford algebra,
\[\Gamma_1=1,\qquad \Gamma_2=\gamma^\mu,\qquad \Gamma_3=\sigma^{\mu\nu},\qquad \Gamma_4=\gamma^\mu\gamma_5,\qquad \Gamma_5=\gamma_5.\]
Then potentials are derived with the full baryon spinor, 
\begin{align}\label{ub}
  u_B(p, s)= N_p
  \left(
  \begin{array}{c}
    1 \\
    \frac{\sigma\cdot p}{E_p+M_B}
  \end{array}\right)
  \chi_s,
  ~~~~~~~~
  N_p=\sqrt{\frac{E_p+M_B}{2M_B}},
\end{align}
where $E_p=\sqrt{p^2+M_B^2}$, and $M_B = M_{Y_c}$ or $M_N$ the mass of charmed baryon or nucleon.
By performing partial wave projection, the $^1S_0$ and $^3S_1-{}^3D_1$ CT potentials in the $LSJ$ basis read
\begin{align}
V_{1S0}^{Y_cN} & = \xi _{{Y_c N}} \left[C_{{1S0}} \left(R_{p'}^N R_{p'}^{{Y_c}}+R_p^N R_p^{{Y_c}}\right)+C'_{{1S0}} \left(R_{p'}^N R_p^N R_{p'}^{{Y_c}} R_p^{{Y_c}}+1\right)\right],\\
V_{3S1}^{Y_cN} & = \frac{1}{9} \xi _{{Y_cN}} \left\{2\left(C_{{1S0}}- C'_{{1S0}} \right)\left( R_{p'}^{{Y_c}} R_p^{{Y_c}}- R_{p'}^N R_p^N\right)  \right.\nonumber\\
& \quad \left.+C_{{3S1}} \left(-6 R_{p'}^N R_p^N+9 R_{p'}^N R_{p'}^{{Y_c}}+9 R_p^N R_p^{{Y_c}}+6 R_{p'}^{{Y_c}} R_p^{{Y_c}}\right)\right.\nonumber\\
& \quad \left.+9 C'_{{3S1}} \left[R_{p'}^{{Y_c}} R_p^{{Y_c}} \left(R_{p'}^N R_p^N-2\right)+2 R_{p'}^N R_p^N+9\right]\right\},\\
V_{3D1-3S1}^{Y_c N} & = \frac{\xi _{{Y_c N}}}{9 \sqrt{2}} \left\{\left(C_{{1S0}} - C'_{{1S0}}\right) \left[R_p^N \left(R_{p'}^N+3 R_{p'}^{{Y_c}}\right)-R_p^{{Y_c}} \left(3 R_{p'}^N+R_{p'}^{{Y_c}}\right)\right]\right.\nonumber\\
& \quad  \left.+C_{{3S1}} \left[9 R_p^{{Y_c}} \left(R_{p'}^N+4 R_p^N\right)+3 R_{p'}^N R_p^N-3 R_{p'}^{{Y_c}} \left(3 R_p^N+R_p^{{Y_c}}\right)\right]\right.\nonumber\\
&\quad \left.+9 C'_{{3S1}} \left\{R_{p'}^{{Y_c}} \left[R_p^N \left(4 R_{p'}^N R_p^{{Y_c}}+3\right)+R_p^{{Y_c}}\right]-R_{p'}^N \left(R_p^N+3 R_p^{{Y_c}}\right)\right\}\right\},\\
V_{3S1-3D1}^{Y_cN} & = \frac{\xi _{{Y_cN}}}{9 \sqrt{2}} \left\{\left(C_{{1S0}}-C'_{{1S0}}\right) \left[R_{p'}^N \left(R_{p}^N+3 R_{p}^{{Y_c}}\right)-R_{p'}^{{Y_c}} \left(3 R_{p}^N+R_{p}^{{Y_c}}\right)\right]\right.\nonumber\\
& \quad  \left.+C_{{3S1}} \left[9 R_{p'}^{{Y_c}} \left(R_{p}^N+4 R_{p'}^N\right)+3 R_{p}^N R_{p'}^N-3 R_{p}^{{Y_c}} \left(3 R_{p'}^N+R_{p'}^{{Y_c}}\right)\right]\right.\nonumber\\
&\quad \left.+9 C'_{{3S1}} \left\{R_{p}^{{Y_c}} \left[R_{p'}^N \left(4 R_{p}^N R_{p'}^{{Y_c}}+3\right)+R_{p'}^{{Y_c}}\right]-R_{p}^N \left(R_{p'}^N+3 R_{p'}^{{Y_c}}\right)\right\}\right\},\\
V_{3D1}^{Y_cN} & = \frac{2}{9} \xi _{{Y_cN}} \left\{\left(C_{{1S0}}-C'_{{1S0}}+3 C_{{3S1}}\right) \left(R_{p'}^N R_p^N-R_{p'}^{{Y_c}} R_p^{{Y_c}}\right) \right.\nonumber\\
& \quad \left.+9 C'_{{3S1}} \left[R_{p'}^N R_p^N \left(4 R_{p'}^{{Y_c}} R_p^{{Y_c}}-1\right)+R_{p'}^{{Y_c}} R_p^{{Y_c}}\right]\right\}.
\end{align}
where 
\begin{align}
{\xi_{Y_cN}} = 4 \pi \frac{\sqrt{\left(E_{p'}^{Y_c}+M_{Y_c}\right)\left(E_{p}^{Y_c}+M_{Y_c}\right)\left(E_{p'}^{N}+M_{N}\right)\left(E_{p}^{N}+M_{N}\right)}}{4M_N M_{Y_c}}~~~~
\textrm{and}
~~~~R_{p(p')}^{Y_c,N}=\frac{p(p')}{E_{p(p')}^{Y_c,N}+M_{Y_c,N}}.
 \end{align} 
Note that all the LECs are implicitly pion mass dependent as in Ref.~\cite{Haidenbauer:2017dua} such that $C_{1S0} = \hat{C}_{1S0} + D_{1S0} ~ m_\pi^2$. For $M_{Y_c}$, we used  the average of $\Lambda_c$ and $\Sigma_c$ masses. On the other hand, because of the limited  LQCD data, it is impossible to pin down all the LECs of the coupled $\Lambda_c N - \Sigma_c N$ system. Therefore, following Ref.~\cite{Erkelenz:1971caz}, we used an effective CT potential by only considering the $\Lambda_c N \rightarrow \Lambda_c N$ channel and assumed that the CT contributions from the $\Sigma_c N$ channel can be effectively absorbed into those from the $\Lambda_c N$ channel, thus in total only four LECs are needed in the present study, i.e. $C_{{1S0}}$, $C'_{{1S0}}$, $C_{{3S1}}$, and $C'_{{3S1}}$.

To construct the OME potentials, we need the following LO meson-baryon Lagrangian~\cite{Yan:1992gz}:
\begin{align}
\begin{split}
 \mathcal{L}_{MB}
=&~\textrm{tr}\left(\bar{B}\left(i\slashed{D}-M_B\right)B - \frac{D/F}{2}\bar{B}\gamma^{\mu}\gamma_5\{u_\mu,B\}_{\pm} \right)\\
~&+ \frac{1}{2}\textrm{tr}(\bar{B}_{\bar{3}}(i\slashed{\partial} -M_{\bar{3}})B_{\bar{3}})+\textrm{tr}\frac{1}{8f_0^2}(i\bar{B}_{\bar{3}}\gamma^\mu\{[M,\partial_\mu M],B_{\bar{3}}\})\\
~&+ \textrm{tr}(\bar{B}_6(i\slashed{\partial}  -M_6)B_{6})+\textrm{tr}\frac{1}{4f_0^2}(i\bar{B}_{6}\gamma^\mu\{[M,\partial_\mu M],B_6\})\\
~&+ (-\frac{1}{\sqrt{2}f_0})g_1\textrm{tr}(\bar{B}_6\gamma^\mu\gamma_5 \partial_\mu MB_6)
+ (-\frac{1}{\sqrt{2}f_0})g_2\textrm{tr}(\bar{B}_6\gamma^\mu\gamma_5 \partial_\mu M B_{\bar{3}})+\textrm{h.c.}\\
~&+ (-\frac{1}{\sqrt{2}f_0})g_6\textrm{tr}(\bar{B}_{\bar{3}}\gamma^\mu\gamma_5 \partial_\mu M B_{\bar{3}}),\\
\end{split}
\end{align}
where tr indicates trace in the corresponding flavor space, $D_\mu B = \partial_\mu B+[\Gamma_\mu,B]$, $\Gamma_\mu$ and $u_\mu$ are the vector and axial vector combinations of the meson fields and their derivatives,
\[
\Gamma_\mu=\frac{1}{2}\left(u^\dag \partial_\mu u + u \partial_\mu u^\dag\right), ~~u_\mu=i(u^\dag \partial_\mu u - u \partial_\mu u^\dag).
\] 
In the Lagrangian $\mathcal{L}_{MB}$, $M_{B}$, $M_{\bar 3}$ and $M_{6}$ are the ground-state masses of octet baryons, antitriplet baryons and sextet baryons, respectively, and $M$, $B$, $B_{\bar 3}$, and  $B_{6}$ refer to the meson and baryon matrices, which are defined as,
\begin{align*}
M =
\left(
  \begin{array}{ccc}
    \frac{\pi^0}{\sqrt{2}}+\frac{\eta}{\sqrt{6}}& \pi^+ & K^+ \\
    \pi^- & \frac{-\pi^0}{\sqrt{2}}+\frac{\eta}{\sqrt{6}} & K^0 \\
    K^- & \bar{K}^0 & -\frac{2\eta}{\sqrt{6}} \\
  \end{array}
\right),
\end{align*}
\begin{align*}
  B =
  \left(
   \begin{array}{ccc}
    \frac{\Sigma^0}{\sqrt{2}}+\frac{\Lambda}{\sqrt{6}} & \Sigma^+ & p \\
    \Sigma^- & -\frac{\Sigma^0}{\sqrt{2}}+\frac{\Lambda}{\sqrt{6}} & n\\
    \Xi^- & \Xi^0 & -\frac{2\Lambda}{\sqrt{6}}
  \end{array}
  \right),~~
B_{\bar{3}}=
\left(
  \begin{array}{ccc}
    0 & \Lambda_c^+ & \Xi_c^+ \\
    -\Lambda_c^+ & 0 & \Xi_c^0  \\
    -\Xi_c^+ & -\Xi_c^0 & 0 \\
  \end{array}
\right),~~
B_6=
\left(
  \begin{array}{ccc}
    \Sigma_c^{++} & \frac{\Sigma_c^+}{\sqrt{2}} & \frac{\Xi_c^{'+}}{\sqrt{2}} \\
    \frac{\Sigma_c^+}{\sqrt{2}} & \Sigma_c^0 & \frac{\Xi_c^{'0}}{\sqrt{2}} \\
    \frac{\Xi_c^{'+}}{\sqrt{2}} &  \frac{\Xi_c^{'0}}{\sqrt{2}} & \Omega_0 .\\
  \end{array}
\right).
\end{align*}
 The values of the coupling constants $g_1$, $g_2$, and $g_6$, and the meson decay constant $f_0$ will be specified below.
Using $\mathcal{L}_{MB}$, one can straightforwardly obtain the OME potential,

\begin{align}\label{OMES}
V^{\rm{OME}}_{Y_cN\rightarrow Y_cN} =&-iN\bar{u}_{Y_c}\left(p'\right)\left(\frac{\gamma^{\mu} \gamma_5 q_{\mu}}{2f_0 }\right)u _{Y_c}(p) \frac{i}{\Delta E^2 - q^2 - m^2+i\epsilon}\nonumber\\ 
~&\times\bar{u}_N\left(-p'\right) \left(\frac{ \gamma^{\nu} \gamma_5 \boldmath{q}_{\nu}}{2f_0}\right) u_N(-p)  \times \mathcal{I}_{Y_cN\rightarrow Y_cN},
\end{align}
where ${Y_c = \Lambda_c, \Sigma_c}$, $q = p^{\prime}-p$ is the transferred momentum, and $m$ is the mass of the exchanged pseudoscalar meson. Note that we only consider light meson exchanges in the ChEFT. The coupling constant $N$ is defined as,
\begin{align}
N = g_{A}^{Y_c Y_c'} g_{A}^{NN}.
\end{align}
Following Ref.~\cite{Haidenbauer:2017dua}, $g_{A}^{Y_c Y_c'}$ and $g_{A}^{NN}$ are assumed to be pion mass independent and are fixed to be $g_{A}^{NN}=1.27$~\cite{Patrignani:2016xqp},  $g_{A}^{\Sigma_c\Sigma_c}=0.71$~\cite{Can:2016ksz}, and $g_{A}^{\Lambda_c\Sigma_c}=0.74$~\cite{Can:2016ksz,Albertus:2005zy}. On the other hand, the meson decay constant $f_0$ varies with the pion mass, and the dependence has been deduced, e.g., in Ref.~\cite{Durr:2013goa}. We use $f_0=93$ MeV for $m_\pi=138$ MeV, $f_0=112$ MeV for $m_\pi=410$ MeV, and $f_0=129$ MeV for $m_\pi=570$ MeV. In Eq.~(\ref{OMES}), $\mathcal{I}$ indicates the isospin factor, whose value can be found in, e.g., Refs.~\cite{Polinder:2006zh, Li:2016paq}. Note that OME does not contribute to $\Lambda_c N \rightarrow \Lambda_c N$ at tree level because of isospin conservation, but it contributes to the scattering amplitudes via the scattering equation.

In order to obtain the scattering amplitudes, we solved the coupled-channel Kadyshevsky equation~\cite{Kadyshevsky:1967rs},
\begin{align}\label{SEK}
  & T_{\rho\rho'}^{\nu\nu',J}(p',p;\sqrt{s})
  =
   V_{\rho\rho'}^{\nu\nu',J}(p',p)
   +
  \sum_{\rho'',\nu''}\int_0^\infty \frac{dp''p''^2}{(2\pi)^3} \frac{M_{B_{1,\nu''}}M_{B_{2,\nu''}}~ V_{\rho\rho''}^{\nu\nu'',J}(p',p'')~
   T_{\rho''\rho'}^{\nu''\nu',J}(p'',p;\sqrt{s})}{E_{1,\nu''}E_{2,\nu''}
  \left(\sqrt{s}-E_{1,\nu''}-E_{2,\nu''}+i\epsilon\right)},
\end{align}
where $\sqrt{s}$ is the total energy of the two-baryon system in the center-of-mass frame and $E_{n,\nu''}=\sqrt{\mbox{\boldmath $p$}''+M_{B_{n,\nu''}}}$, $(n=1,2)$. The labels $\nu,\nu',\nu''$ denote the particle channels, and $\rho,\rho',\rho''$ denote the partial waves. In numerical calculations, the potentials in the scattering equation are regularized with an exponential form factor of the following form,
\begin{align}\label{EF}
  f_{\Lambda_F}(p,p') = \exp \left[-\left(\frac{p}{\Lambda_F}\right)^{4}-\left(\frac{p'}{\Lambda_F}\right)^{4}\right].
\end{align}
More details about the covariant ChEFT can be found in Refs.~\cite{Ren:2016jna,Li:2016mln,Song:2018qqm,Ren:2018xxd,Li:2018tbt,Bai:2020yml}.

\section{Results and discussion}

\subsection{Fitted results and extrapolations to the physical point}

\begin{figure}[h]
  \centering
  \includegraphics[width=0.85\textwidth]{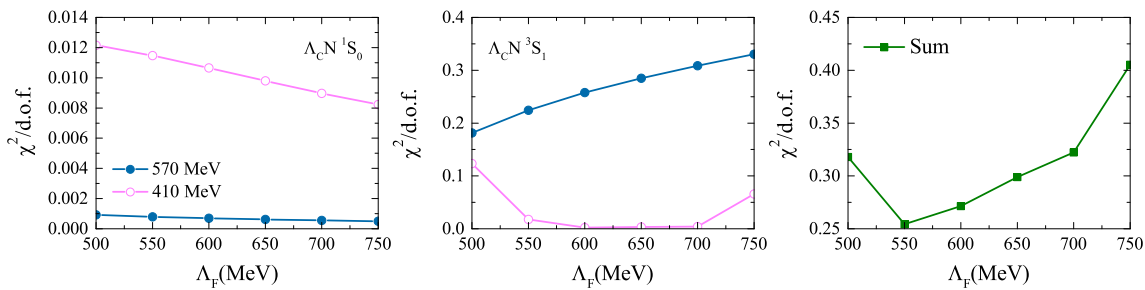}
  \caption{Best fitted $\chi^2/\textrm{d.o.f.}$ as a function of the cutoff in the LO covariant ChEFT by fitting to the lattice QCD $\Lambda_cN$ $S$-wave phase shifts. The magenta circles denote the $\chi^2/\textrm{d.o.f.}$ for $m_\pi = 410$~MeV, and the dark-blue dots refer to the $\chi^2/\textrm{d.o.f.}$ for $m_\pi = 570$~MeV, while their sum is  shown in the right panel.}\label{Chi2}
\end{figure}

The four LECs in the CT potential are determined by fitting to the lattice QCD simulations from the HAL QCD Collaboration~\cite{Miyamoto:2017tjs}. For this, we used the $\Lambda_c N$ $S$-wave phase shifts for $m_\pi = 410$ MeV and $570$ MeV with the center-of-mass energy $E_{\textrm{c.m.}}$ up to $30$ MeV. Although the lattice QCD results for $m_\pi=570$ MeV are probably already beyond the applicability of leading order ChEFT, we included these results in order to pin down the pion mass dependence of the LECs so that we can extrapolate the lattice QCD results to the physical point. In Table~\ref{masses}, we list the lattice QCD ~\cite{Miyamoto:2017tjs} and physical~\cite{Patrignani:2016xqp} baryon masses relevant to the present study.
\begin{table}[h]
\centering
 \caption{Baryon masses for different pion masses (in units of MeV) needed in this work~\cite{Miyamoto:2017tjs}.}~\label{masses}
  \begin{tabular}{L{1cm}C{3cm}C{3cm}R{2cm}}
  \hline
  \hline
$m_\pi$ & $m_N$ & $m_{\Lambda_c}$ & $m_{\Sigma_c}$  \\
\hline
$ 138 $ & $\phantom{0}939$ & $2287$ & $2455$ \\
$412$  & $1215$ & $2434$ & $2575$ \\
$570$  & $1399$ & $2555$ & $2674$\\
\hline
 \hline
\end{tabular}
 \end{table}  
Similar to our previous study on the $\Lambda N-\Sigma N$ system~\cite{Li:2016mln,Song:2018qqm,Li:2018tbt}, the fits were first performed with cutoff values in the range of $\Lambda_F=500-750$~MeV. The fitted $\chi^2$ for lattice QCD simulations for different pion masses and the total $\chi^2$ as a function of the cutoff $\Lambda_F$ are shown in Fig.~\ref{Chi2}. For the $^1S_0$ channel, within the cutoff range studied, the $\chi^2/\mathrm{d.o.f.}$ for $m_\pi=410$ MeV decreases with increasing $\Lambda_F$, while the $\chi^2/\mathrm{d.o.f.}$ for $m_\pi=570$ MeV stays almost constant. On the the other hand, for the $^3S_1$ channel, the $\chi^2/\mathrm{d.o.f.}$ increases with increasing $\Lambda_F$ for $m_\pi=570$ MeV, but stabilizes between $\Lambda_F=550\sim700$ MeV form $m_\pi=410$ MeV. From the right panel, one can see that a cutoff between 500 and 700~MeV yields the minimum $\chi^2$ for all the lattice QCD data fitted. However, since the lattice QCD data for $m_\pi=570$~MeV were used in our fits, in principle the cutoff value should be larger than the pion mass such that the inclusion of the OME potential is justified. As a result, we choose the range of $\Lambda_F=600-700$ MeV in subsequent analyses, and the fitted LECs are shown in Table~\ref{LECs}.
 
\begin{table}[H]
\centering
\caption{Values of the LECs from the best fits obtained with  $\Lambda_F=600-700$ MeV, where $C'_{{1S0}}$ and $C'_{{3S1}}$ are in units of $10^4~\textrm{GeV}^{-2}$, and $C_{{1S0}}$ and $C_{{3S1}}$ are in units of $10^2~\textrm{GeV}^{-2}$. The pion mass $m_\pi$ and cutoff $\Lambda_\textrm{F}$ are in units of MeV.}\label{LECs}
\setlength{\tabcolsep}{20pt}
\begin{tabular}{cccccc}
 \hline
 \hline
$m_\pi$& $\Lambda_\textrm{F}$& $C_{{1S0}}$ & $C'_{{1S0}}$ & $C_{{3S1}}$ & $C'_{{3S1}}$ \\
 \hline
\multirow{3}*{$410$} & $600$ & $-1.2653$ & $1.6882$ & $\phantom{-}0.2698$ & $1.8966$   \\
                     & $650$ & $-0.9267$ & $1.8736$ & $\phantom{-}0.1270$ & $1.3170$   \\
                     & $700$ & $-0.2255$ & $2.1488$ & $-0.0130$ & $0.8456$   \\
\hline
\multirow{3}*{$570$} & $600$ & $-0.7624$ &  $0.6540$ & $-0.0520$ & $0.1994$   \\
                     & $650$ & $-0.7168$ &  $0.6876$ & $-0.0468$ & $0.1608$   \\
                     & $700$ & $-0.6485$ &  $0.7274$ & $-0.0414$ & $0.1323$   \\
 \hline
 \hline
\end{tabular}
\end{table}

\begin{figure}[h]
  \centering
  \includegraphics[width=0.35\textwidth]{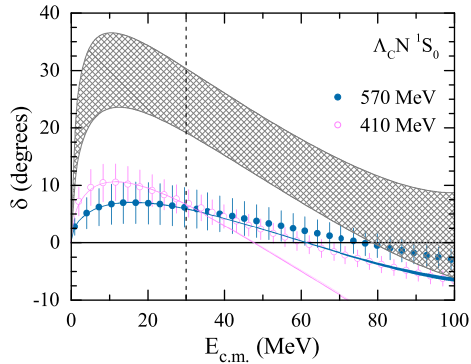} 
  \caption{$\Lambda_cN$ $^1S_0$ phase shifts of the lattice QCD simulations in comparison with the ChEFT fits. The magenta opened circles denote the LQCD data~\cite{Miyamoto:2017tjs} for $m_\pi=410$~MeV, while the dark-blue dots refer to the LQCD data~\cite{Miyamoto:2017tjs} for $m_\pi=570$~MeV. The lines/bands denote the ChEFT fits. The bands are generated from a variation of $\Lambda_\textrm{F}$ from $600$~MeV to $700$~MeV. The grey lines/bands refer to the predictions for $m_\pi=138$~MeV. The vertical dashed lines at $E_\mathrm{c.m.}$ denote that the $\Lambda_c N$ interaction is obtained by fitting to the lattice QCD data only up to this energy.}\label{1s0_mass}
\end{figure}

\begin{figure}[h]
  \centering
  \includegraphics[width=0.35\textwidth]{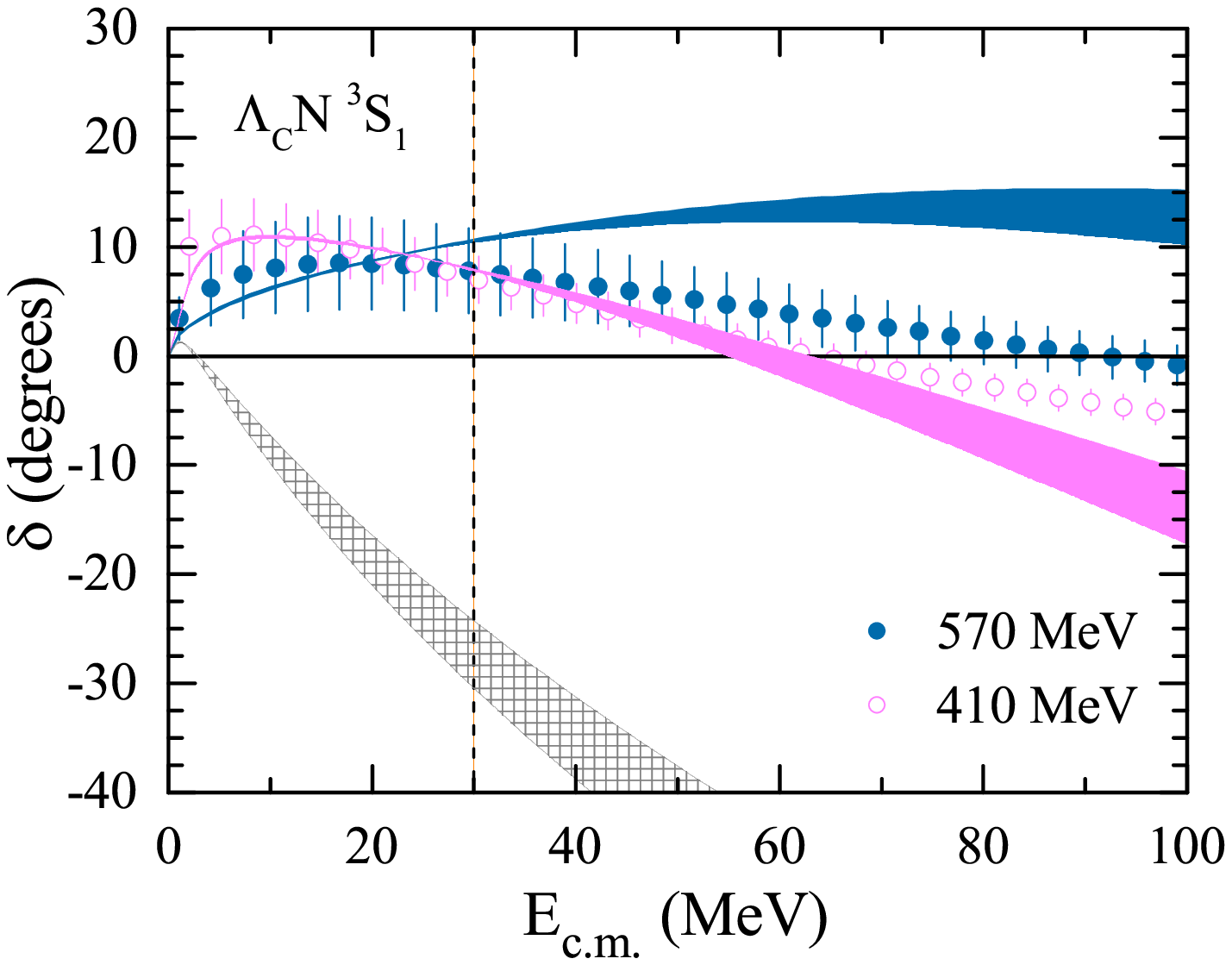}\qquad
  \includegraphics[width=0.35\textwidth]{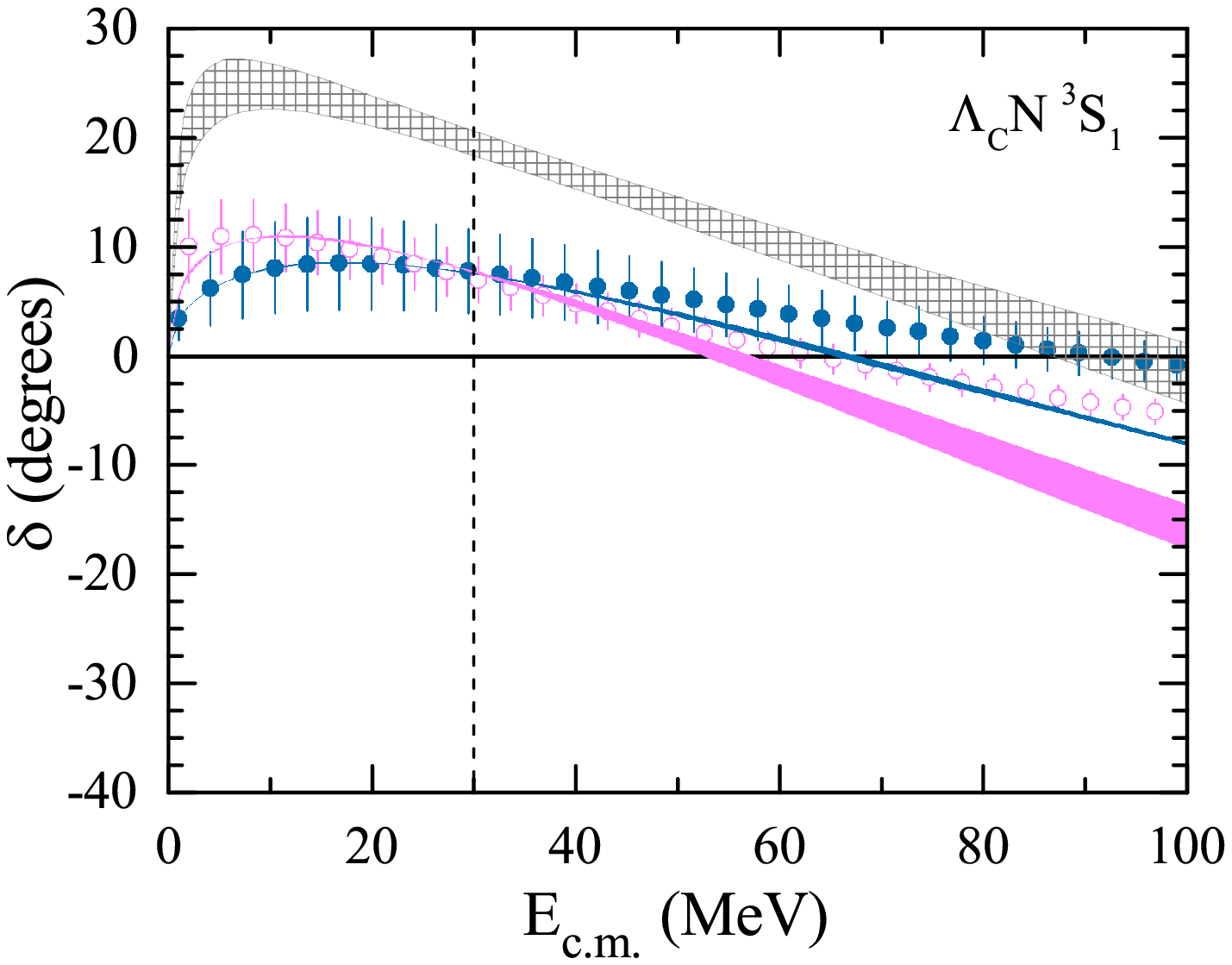}
  \caption{Left: the same as Fig.~\ref{1s0_mass} but for the $^3S_1$ phase shifts. Right: the same as the left panel, but with the $S-D$ coupled channel effect turned off. The vertical dashed lines at $E_\mathrm{c.m.}$ denote that the $\Lambda_c N$ interaction is obtained by fitting to the lattice QCD data only up to this energy.}\label{CLOSE-3DS1}
\end{figure}

The fitted $S$-wave $\Lambda_cN$ phase shifts are shown in  Figs.~\ref{1s0_mass} and ~\ref{CLOSE-3DS1} . For the $^1S_0$ partial wave, the covariant ChEFT phase shifts are in good agreement with the lattice QCD data. The $\Lambda_cN$ potential turns out to be moderately attractive when extrapolated to the physical point. It should be noted that the predicted $^1S_0$ phase shifts at the physical point are qualitatively similar to those of the non-relativistic ChEFT of Ref.~\cite{Haidenbauer:2017dua}, but with slightly larger uncertainties.

On the other hand, for the $^3S_1$ partial wave shown in the left panel of  Fig.~\ref{CLOSE-3DS1}, the covariant ChEFT phase shifts are in fair agreement with the lattice QCD data only for energies up to $30$ MeV. The discrepancy then becomes larger as  the energy increases. Extrapolated to the physical point, the $\Lambda_cN$ interaction is weakly attractive only at the  very low energy region then becomes repulsive as the kinetic energy increases. This results in a peculiar phenomenon that although the scattering length of this channel is negative (see Table~\ref{SL}) and therefore indicates a weakly attractive interaction, on the whole the $^3S_1$ interaction is repulsive. This is quite different from the results of the non-relativistic ChEFT~\cite{Haidenbauer:2017dua}. 

In order to understand this phenomenon, we set $V_{^3S_1-^3D_1}$ and $V_{^3D_1-^3S_1}$ in the CT potential to zero, and redid the fits. The resulting phase shifts are shown in the right panel of Fig.~\ref{CLOSE-3DS1}. The covariant ChEFT phase shifts are in better agreement with the lattice QCD data for energies up to $40$ MeV. In particular, the description of the $m_\pi=570$ MeV data is much improved. Furthermore, when extrapolated to the physical point, an attractive interaction is obtained, which is similar not only to the $^1S_0$ interaction shown in Fig.~\ref{1s0_mass}, but also to the $^3S_1$ interaction of the non-relativistic ChEFT. As a result, we conclude that the predicted $^3S_1$ interaction depends strongly on how the coupled channel $S-D$ mixing is treated. 

In the non-relativistic ChEFT, there is no $S-D$ mixing in the leading order CT potential, while the same two LECs are responsible for the $^3S_1-{}^3D_1$ coupled channel in the LO covariant ChEFT. In the following, we further explore this coupled channel effect which is due to relativistic corrections that are considered as of higher order in the non-relativistic ChEFT but already shows up at leading order in the covariant ChEFT.

\begin{figure}[h]
  \centering
  \includegraphics[width=0.85\textwidth]{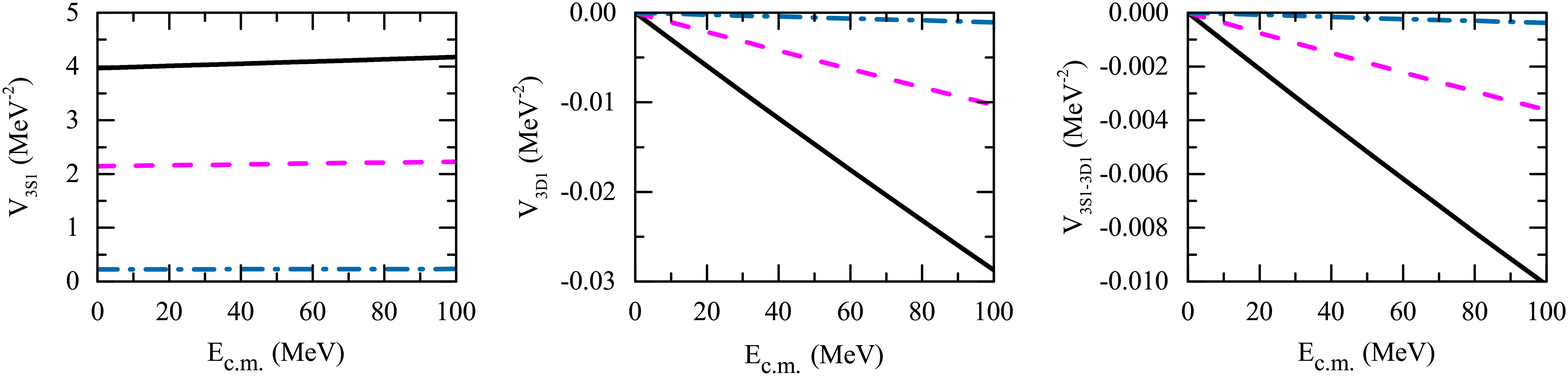}
  \caption{On-shell $\Lambda_cN$ potentials for the $^3S_1-{}^3D_1$ coupled channel for different pion masses. The black solid lines denote potentials for $m_\pi$= $138~\textrm{MeV}$, the magenta dashed lines refer to potentials for $m_\pi$= $410~\textrm{MeV}$,  and the dark-blue dotted lines are potentials for $m_\pi$= $570~\textrm{MeV}$.}\label{V3DS1}
\end{figure}

For the sake of simplicity, we fixed the cutoff $\Lambda_F$ at $600$ MeV, but the general discussion remain unchanged for
$\Lambda_F=650$ and 700 MeV. The on-shell coupled channel potentials as a function of kinetic energy in the center of mass frame are shown in ~Fig.\ref{V3DS1}. The $^3S_1$ potential increases slowly  with  $E_\textrm{c.m.}$, while the $^3D_1$ and $^3S_1-{}^3D_1$ potentials decrease with the energy. The $^3D_1$ potential is two order of magnitude smaller than the $^3S_1$ potential, while the $^3S_1-{}^3D_1$ mixing is even one more order of magnitude smaller. Such features of the triplet channel potentials are in qualitative agreement with the lattice QCD simulations~\cite{Miyamoto:2017tjs}. Nonetheless, the small mixing seems to affect the $^3S_1$ phase shifts a lot, as we noted above. As a result, we strongly encourage lattice QCD collaborations to check whether the inclusion of the coupled channel effect in extrapolating the $^3S_1$ interaction can make a difference.  

\begin{figure}[h]
  \centering
  \includegraphics[width=0.85\textwidth]{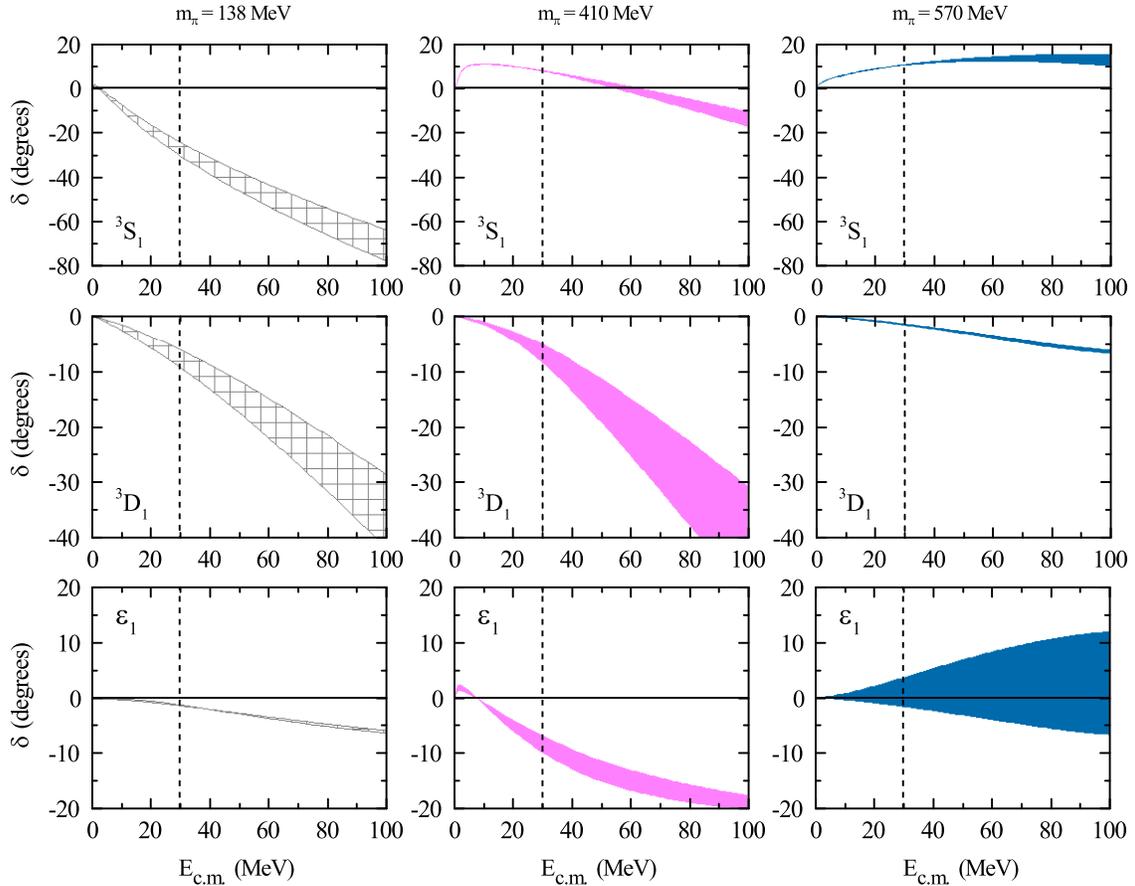}
  \caption{$\Lambda_c N$ $^3S_1$, $^3D_1$ phase shifts and mixing angles $\varepsilon_1$ for different pion masses. The bands are generated from the variation of $\Lambda_F$ from $600$ MeV to $700$ MeV. The vertical dashed lines at $E_\mathrm{c.m.}$ denote that the $\Lambda_c N$ interaction is obtained by fitting to the lattice QCD data only up to this energy.}\label{mixing_angle_bands}
\end{figure}

In the following, we predict the $^3D_1$ phase shifts and the mixing angle $\varepsilon_1$ with the LECs determined by fitting to the lattice QCD $^3S_1$ phase shifts, which are shown in Fig.~\ref{mixing_angle_bands} together with their $^3S_1$ counterparts for three pion masses, $m_\pi=138$, $410$, and $570$ MeV.  As this is only a leading order study and also demonstrated above, the results beyond $E_\mathrm{c.m.}>30$ MeV should be taken with caution. For the $^3S_1$ partial wave, the interaction changes from weakly attractive to moderately repulsive as the pion mass decreases from $570$ MeV to $138$ MeV. As already stressed above, this transition is strongly related to the $S-D$ coupling and should be checked by future lattice QCD simulations.  For the $^3D_1$ partial wave, the interaction is weakly repulsive for $m_\pi=570$ MeV, and becomes stronger as the pion mass decreases to $m_\pi=410$ MeV. However, such a reduction does not extend to the physical point. As a matter of fact, the $\Lambda_cN$ interaction for the physical pion mass is almost the same as that for $m_\pi=410$ MeV. As for the mixing angle $\varepsilon_1$, it shows a strong dependence on the cutoff for $m_\pi= 570$ MeV, yet such a dependence becomes weaker as the pion mass decreases. In addition, for $m_\pi=410$ MeV, the mixing angle has the largest magnitude.

\subsection{Comparison of the  $NN$, $\Lambda N$ and $\Lambda_c N$ interactions}
It is instructive to compare the $NN$, $\Lambda N$ and $\Lambda_c N$ interactions, which not only allows for a better understanding of the evolution of the baryon-nucleon interactions as one replaces one light quark in the baryon by one strange or charm quark, but also allows us to better assess the extrapolated $\Lambda_c N$ interaction from the lattice QCD simulations as both the $NN$ and $YN$ interactions are constrained by experimental data, particularly the former.

\begin{figure}[h]
  \centering
  \includegraphics[width=0.7\textwidth]{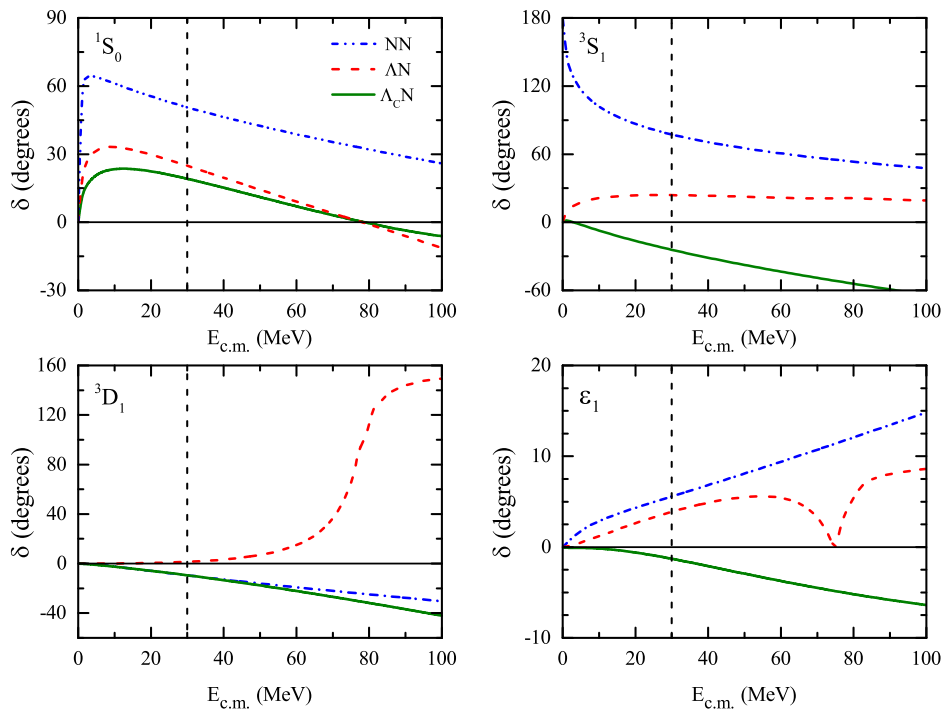}
  \caption{$NN$, $\Lambda N$ and $\Lambda_c N$ phase shifts in the $^1S_0$ and $^3S_1-{}^3D_1$ coupled channels for physical pion masses. The  green solid lines, red dashed lines, and blule dot-dashed lines denote the $\Lambda_c N$, $\Lambda N$, and $NN$ interactions, respectively. The vertical dashed lines at $E_\mathrm{c.m.}$ denote that the $\Lambda_c N$ interaction is obtained by fitting to the lattice QCD data only up to this energy.}\label{quark_dependences}
\end{figure}

The $^1S_0$ and $^3S_1-{}^3D_1$ $NN$, $\Lambda N$ and $\Lambda_c N$ phase shifts with $\Lambda_\textrm{F}=600$ MeV for the physical pion mass are shown in Fig.~\ref{quark_dependences}. In the $^1S_0$ channel, the $NN$ interaction is strongly attractive (to the extent that there is a virtual bound state in this channel), while both the $\Lambda N$ and $\Lambda_c N$ interactions become less attractive, but the latter two are of similar strength. In the $^3S_1$ channel, the $NN$ interaction is strongly attractive (to the extent that the deuteron exists), the $\Lambda N$ interaction becomes only weakly attractive, while the $\Lambda_c N$ interaction becomes weakly repulsive, as we discussed already. In the $^3D_1$ channel, both the $NN$ and $\Lambda_c N$ interactions are weakly repulsive, while the $\Lambda N$ interaction is quite small up to $E_\mathrm{c.m.}\sim50$ MeV, and then increases quickly, corresponding to the opening of the $\Sigma N$ channel. Such a phenomenon has been observed in the experimental data~\cite{Kadyk:1971tc,Hauptman:1977hr} and our previous studies~\cite{Li:2016mln,Li:2016paq} that a cusp appears in the $\Lambda p \rightarrow \Lambda p$ cross section at the opening of the $\Sigma^+n$ channel. On the other hand, the $NN$ and $\Lambda N$ mixing angles are very similar up to $E_\mathrm{c.m.}\sim 50$ MeV but the $\Lambda_c N$ mixing angle is negative and smaller in magnitude.   

In Table~\ref{SL}, we list the corresponding scattering lengths. For comparison, we also show the results of the LO covaraint ChEFT~\cite{Li:2016mln} for $\Lambda N$, next-to-leading order non-relativistic ChEFT~\cite{Haidenbauer:2019boi,Haidenbauer:2017dua} for $\Lambda N$ and $\Lambda_c N$, and Nijmegen-D model~\cite{Gibson:1983zw,Nagels:1976xq}.   The covariant ChEFT $NN$ scattering lengths were obtained with the  regulator of Eq.~(\ref{EF}) with a cutoff of 600 MeV following the fitting strategy of Ref.~\cite{Ren:2016jna} while those of the non-relativistic ChEFT were obtained  using the same regulator and cutoff following  the fitting strategy of Ref.~\cite{Epelbaum:1999dj}.   We note that the scattering lengths obtained in the covariant ChEFT are qualitatively similar to those obtained in the non-relativistic ChEFT, while some of the predictions of the Nijmegen-D model are drastically different, e.g., those of the $\Lambda_c N$ and $^1S_0$ $NN$. 

\begin{table}[h]
\centering
 \caption{$NN$, $\Lambda N$ and $\Lambda_c N$ scattering lengths (in units of fm)  obtained in the LO covariant ChEFT, NLO non-relativistic ChEFT, and Nijmegen-D model. For guidance, we also show the experimental $NN$ scattering lengths. }\label{SL}
\setlength{\tabcolsep}{14.5       pt}
 \begin{tabular}{cccccc}
  \hline
  \hline
  Channels &~& Cov. ChEFT (LO) &  NR ChEFT (NLO) & Nijmegen-D & Exp.\\
  \hline
\multirow{2}{*}{$NN$} & $a_{1S0}^{NN}$ & $-21.3$\phantom{~\cite{Li:2016mln}} & $-23.0$\phantom{~\cite{Haidenbauer:2019boi}}  & $-17.0 $~\cite{Gibson:1983zw}  & $-23.7$\\
 & $a_{3S1}^{NN}$ & $\phantom{-}5.75$\phantom{~\cite{Li:2016mln}} & $\phantom{-}5.48$\phantom{~\cite{Haidenbauer:2019boi}}  & $ \phantom{-}5.42$~\cite{Gibson:1983zw} & $\phantom{-}5.42$\\
\hline
\multirow{2}{*}{$\Lambda N$} & $a_{1S0}^{\Lambda N}$ & $-2.44$~\cite{Li:2016mln} & $-2.91$~\cite{Haidenbauer:2019boi} & $-1.90$~\cite{Nagels:1976xq}  \\
 & $a_{3S1}^{\Lambda N}$ & $-1.32$~\cite{Li:2016mln} & $-1.54$~\cite{Haidenbauer:2019boi} &  $-1.96$~\cite{Nagels:1976xq} \\
\hline
\multirow{2}{*}{$\Lambda_c N$} & $a_{1S0}^{\Lambda_c N}$ & $-1.16$\phantom{~\cite{Li:2016mln}} & $-1.00$~\cite{Haidenbauer:2017dua}  & $-3.83$~\cite{Gibson:1983zw}  \\
  & $a_{3S1}^{\Lambda_c N}$ & $-0.52$\phantom{~\cite{Li:2016mln}} & $-0.98$~\cite{Haidenbauer:2017dua} & $-4.24$~\cite{Gibson:1983zw}  \\
 \hline
 \hline
\end{tabular}
 \end{table}

\section{Conclusion}

In this work, we studied the lattice QCD simulations of the $\Lambda_c N$ interaction for $m_\pi = 410, 570$ MeV and extrapolated the results to the physical point. We found that the covariant ChEFT $\Lambda_c N$ phase shfits are in good agreement with the lattice QCD simulations in the $^1S_0$ partial wave for $E_{\text{c.m.}}\leq 40$ MeV, while in the $^3S_1$ partial wave, the ChEFT results are in good agreement with the lattice QCD data for $E_{\text{c.m.}}\leq 30$ MeV. Different from the previous study of Ref.~\cite{Haidenbauer:2017dua}, we found an attractive $\Lambda_c N$ interaction in the $^1S_0$ partial wave, but a repulsive interaction in the $^3S_1$ partial wave, though it is slightly attractive at extremely low energies. We showed that such a repulsive $\Lambda_c N$ interaction is originated from the coupling of $^3S_1$ and $^3D_1$ and one ends up with an attractive interaction at the physical point once the coupling in the CT potential is neglected. 

To understand how the baryon-nucleon ($BN$) interaction evolves if one replaces one of the light quarks in the baryon $B$
 with a strange or charm quark, We  compared the so-obtained $\Lambda_c N$ interaction with those of $\Lambda N$ and $NN$.   We found that in general the strength of the $BN$ interaction becomes weaker as one moves from $NN$ to $\Lambda N$ to $\Lambda_c N$.

We noted that although in good agreement with the lattice QCD simulations for unphysical pion masses, the covariant ChEFT results for the spin triplet channel seem to show strong dependence on the consideration of coupled channel effects. This needs to be checked by future and more refined lattice QCD simulations.

\section{Acknowledgements}

This work was partly supported by the National Natural Science Foundation of China (NSFC) under Grants No. 11975041, No.11735003, and No.11961141004. Yang Xiao acknowledges the support from China Scholarship Council.

\end{document}